\documentclass[english,aps,prx,notitlepage,onecolumn,showpacs,superscriptaddress,,nofootinbibfloatfix]{revtex4-1}
\usepackage[T1]{fontenc}
\usepackage[latin1]{inputenc}
\usepackage{graphicx}
\usepackage{bm}
\usepackage{amssymb}
\usepackage{color}
\usepackage[usenames,dvipsnames]{xcolor}
\usepackage{amsmath}
\usepackage{amstext}
\usepackage{latexsym}
\usepackage[colorlinks=true,citecolor=Cerulean,linkcolor=RubineRed,urlcolor=Cerulean]{hyperref}

\usepackage{mathptmx} 
\usepackage{verbatim}

\usepackage{amsfonts}
\usepackage{epsfig}
\usepackage{dsfont}
\usepackage{mathrsfs}
\usepackage{arydshln,leftidx,mathtools}

\begin{document}

\title{Scaling-up quantum heat engines efficiently via shortcuts to adiabaticity}

\author{M. Beau}
\affiliation{Department of Physics, University of Massachusetts, Boston, MA 02125, USA}
\affiliation{Dublin Institute for Advanced Studies, School of Theoretical Physics, Dublin 4, Ireland}
\author{J. Jaramillo}
\affiliation{Department of Physics, National University of Singapore, Singapore 117551}
\affiliation{Department of Physics, University of Massachusetts, Boston, MA 02125, USA}
\author{A. del Campo}
\affiliation{Department of Physics, University of Massachusetts, Boston, MA 02125, USA}

\def\q{{\bf q}}

\def\G{\Gamma}
\def\L{\Lambda}
\def\la{\lambda}
\def\g{\gamma}
\def\al{\alpha}
\def\s{\sigma}
\def\e{\epsilon}
\def\k{\kappa}
\def\ve{\varepsilon}
\def\l{\left}
\def\r{\right}
\def\te{\mbox{e}}
\def\d{{\rm d}}
\def\t{{\rm t}}
\def\K{{\rm K}}
\def\N{{\rm N}}
\def\H{{\rm H}}
\def\la{\langle}
\def\ra{\rangle}
\def\om{\omega}
\def\Om{\Omega}
\def\vep{\varepsilon}
\def\wh{\widehat}
\def\tr{{\rm Tr}}
\def\da{\dagger}
\def\iz{\left}
\def\zi{\right}
\newcommand{\beq}{\begin{equation}}
\newcommand{\eeq}{\end{equation}}
\newcommand{\beqa}{\begin{eqnarray}}
\newcommand{\eeqa}{\end{eqnarray}}
\newcommand{\intf}{\int_{-\infty}^\infty}
\newcommand{\into}{\int_0^\infty}

\begin{abstract}
The finite-time operation of a quantum heat engine that uses a single particle as a working medium generally increases the output power at the expense of inducing friction that lowers the cycle efficiency. 
We propose to scale up  a quantum heat engine utilizing a many-particle working medium in combination with the use  of shortcuts to adiabaticity to boost the nonadiabatic performance by eliminating quantum friction and reducing the cycle time.
To this end, we first analyze the finite-time thermodynamics of a quantum Otto cycle implemented with a  quantum fluid confined in a time-dependent harmonic trap. 
We show that nonadiabatic effects can be controlled and tailored  to match the adiabatic performance using a variety of shortcuts to adiabaticity. 
As a result, the nonadiabatic dynamics of the scaled-up many-particle quantum heat engine exhibits no friction and the cycle can be run at maximum efficiency with a tunable output power.
We demonstrate our results with a working medium consisting of particles with inverse-square pairwise interactions, that includes noninteracting and hard-core bosons as limiting cases. 
\end{abstract}

\maketitle

\section{Introduction}

Quantum thermodynamics resembles a fruitful crucible of research fields where the foundations of physics, information science and statistical mechanics merge \cite{bookQThermo04,VA15}.
It is further spurred by the development of quantum technologies that have facilitated the realization and control of thermal machines and related devices exhibiting quantum dynamics.
A prominent example is that of quantum heat engines (QHE) that transform thermal energy into mechanical work.

In both classical and quantum domains, the performance of heat engines is characterized by the efficiency and power of the cycle.
Studies to date have been limited to the optimization of thermodynamic cycles operating with a single-particle working medium.
After the pioneering works \scalebox{.95}[1.0]{\cite{Alicki79,Kosloff84}}, the quantum Otto cycle has been analyzed in detail \cite{Quan1,Abah14,Zhang14,Deng13,delcampo14,Stefanatos14,ZP15,Abah16}. 
It is of relevance to current experimental efforts aimed at realizing a QHE with a single trapped ion \cite{Abah12,Rossnagel14,Singer16}. 
Nonetheless, it is worth pointing out that a universal behavior emerges among different types of cycles in the limit of small action per cycle \cite{Uzdin15}.
These works show that when a single-particle QHE is operated in a finite amount of time, nonadiabatic excitations act as quantum friction, reducing the efficiency of the engine.
As a result, the maximum efficiency is achieved for long cycle times, in the adiabatic limit, when the output power of the QHE vanishes.
This state of affairs is already present in classical heat engines and gave rise to the field of finite-time thermodynamics.

Scaling up a QHE arises as a natural strategy to compensate the tradeoff between efficiency and power. 
Yet, the analysis of QHE with a many-particle working medium remains essentially unexplored. Only recently, a few results have been reported \cite{Kim11,MD14,Campisi15,ZP15,JBdC15,CF16}. 
In particular, the efficiency of an adiabatic QHE operating an Otto cycle with non-interacting bosons and fermions has been studied in \cite{ZP15}. 
It was shown that signatures of quantum statistics are only relevant when the working medium is confined in an anharmonic trap, with an inhomogeneous energy level spacing. 
More recently, we analyzed a nonadiabatic quantum Otto cycle with an infinite family of quantum fluids as a working medium. 
Under harmonic confinement, it was shown that nonadiabatic many-particle quantum effects can enhance the efficiency of the cycle operating 
at maximum output power by up to 50$\%$ of the single-particle counterpart \cite{JBdC15}. 
This enhancement was demonstrated in the limit of very fast driving, with a sudden trap frequency change accounting for the compression and expansion strokes.
Further efforts in many-particle thermodynamics have shown that the performance of a heat engine can be boosted when the working substance is in the vicinity of a phase transition, 
an effect that holds in the classical case \cite{CF16}.

An alternative approach to boost the performance of QHE resorts on the use of shortcuts to adiabaticity (STA) that mimic adiabatic dynamics without the requirement of slow driving \cite{Torrontegui10}. 
STA constitutes a disruptive paradigm in finite-time thermodynamics that avoids the need to sacrifice efficiency for power. This is achieved by tailoring excitations during the nonequilibrium process. 
STA applicable to the expansion and compression strokes often used in thermodynamic cycles has been extensively studied in recent years, both theoretically \cite{Salamon09,Rezek09,Chen10,MN10,Stefanatos10,delcampo11,Schaff11,Hoffmann11,Choi11,Choi12,Choi13,Jarzynski13,OM14} and experimentally with ultracold atomic gases \cite{Schaff1,Schaff2}.
The development of STA applicable to many-body systems has been reported in \cite{delcampo11b,Choi11,Schaff11,DB12,DRZ12,delcampo13,DJD14,Saberi14} and further demonstrated in the laboratory \cite{Rohringer15}.
At the single-particle level, the application of STA to quantum thermodynamics was pioneered in \cite{Deng13,delcampo14}. 
It was shown that STA can assist the finite-time operation of a QHE with a single trapped particle as a working medium to achieve a tunable output power and zero friction. Similar applications have been discussed in quantum refrigerators \cite{Rezek09}.

In this manuscript, we combine both strategies to engineer a scaled-up many-particle QHE whose operation is assisted by STA to boost its performance, maximizing its efficiency and enhancing the output power. 
In particular, we analyze the finite-time thermodynamics of a QHE with a quantum fluid as a working medium. 
For the latter, we consider an infinite family of scale-invariant many-body systems confined in a time-dependent harmonic trap.
We show that the nonequilibrium dynamics of the many-particle thermodynamic cycle can be engineered via STA to run the QHE with zero friction. 
The enhancement of the output power arises from the many-particle nature of the working medium and the reduced cycle time.
The resulting many-particle QHE operates at the maximum possible efficiency, which is shared by classical and quantum heat engines as an upper bound.

\section{Modeling a Many-Particle Quantum Heat Engine}

\subsection{Trapped Quantum Fluids as Working Media}

We introduce a generalization of the single-particle QHE to the many-particle case in which the working medium consists of a quantum fluid that is confined in a time-dependent harmonic trap~\cite{JBdC15}. 
In particular, we focus on a broad family of quantum many-body systems described by \mbox{the Hamiltonian:}
 \begin{equation}
\hat{H} =\sum_{i=1}^{\N}\left[-\frac{\hbar^2}{2m}\nabla^2+\frac{1}{2}m\omega(t)^2 {\bf r}_{i}^{2}\right]+
\sum_{i<j}V({\bf r}_i-{\bf r}_j),
 \label{Hsu11}
\end{equation}
 where $\N$ is the total number of particles and $\omega(t)$ is the trap frequency of the isotropic harmonic confinement. 
 The only condition imposed on the pairwise interaction is that it is a homogeneous function of degree $-2$, \emph{i.e.},
 $V(b{\bf r})=b^{-2}V({\bf r})$. Therefore, Hamiltonian (\ref{Hsu11}) accounts for a wide variety of many-body systems that include free non-interacting gases, quantum fluids with hard-core interactions, inverse-square interactions models and Bose gases with s-wave contact interactions in two spatial dimensions ($d=2$), 
 among other prominent examples.
We note that under a modulation of the trapping frequency $\om(t)$,
the unitary dynamics generated by Equation (\ref{Hsu11}) is governed by scale invariance~\cite{Gambardella75}. As a result, the time-evolution of a many-particle eigenstate $\Psi$ with energy $\varepsilon$ at $t=0$ reads \cite{delcampo11b,JBdC15}: 
\beqa
\Psi({\bf r}_{1},\dots,{\bf r}_{\N},t)=b^{-\frac{\N d}{2}}\exp\left(i\sum_j\frac{m\dot{b}{\bf r}_{j}^{2}}{2\hbar b}-i\frac{\varepsilon}{\hbar}\int_0^t\frac{dt'}{b(t')^2}\right)
\Psi\left(\frac{{\bf r}_{1}}{b},\dots,\frac{{\bf r}_{\N}}{b},t=0\right),
\eeqa
where the positive-definite function of time $b=b(t)>0$ is known as the scaling factor and fulfills the Ermakov differential equation:
\beqa
\label{Ermakoveq}
\ddot{b}+\om^2(t)b=\om^2(0)/b^{3}.
\eeqa

To describe the many-particle QHE, we shall consider an initial equilibrium state in the canonical ensemble at (inverse) temperature $\beta$ described by the density matrix:
\beqa
\hat{\rho}=\frac{e^{-\beta \hat{H}}}{\tr(e^{-\beta \hat{H}})}.
\eeqa

Provided that at $t=0$, the state is thermal, the nonadiabatic mean-energy following a variation of the trapping potential is given by the scaling law \cite{JBdC15}:
\beqa
\label{aven0}
\la \hat{H}(t)\ra_\beta=\frac{Q^{\ast}(t)}{b_{\rm ad}^2}\la \hat{H}(0)\ra_\beta\ ,
\eeqa
where $\la \hat{H}(0)\ra_\beta$ is the mean energy of the initial thermal state, and
 the nonadiabatic factor $Q^{\ast}\geq~1$~reads:
\beqa
\label{Qstar}
Q^{\ast}(t)&=&b_{\rm ad}^2\left(\frac{1}{2b^2}+\frac{\om(t)^2}{2\om_0^2}b^2+\frac{\dot{b}^2}{2\om_0^2}\right).
\eeqa

Here, $b(t)$ is the solution of Equation (\ref{Ermakoveq}) subject to the boundary conditions $b(0)=1$ and \mbox{$\dot{b}(0)=0$,} to account for the initial equilibrium state. We note that in the adiabatic limit $\ddot{b}\approx 0$ and: 
\beqa
b(t)&\rightarrow& b_{\rm ad}=[\om(0)/\om(t)]^{1/2},\\
Q^{\ast}(t)&\rightarrow& Q_{\rm ad}^{\ast}(t)=1.
\eeqa

Since the mean value of the energy following an adiabatic protocol is precisely given by $\la \hat{H}(t)\ra_{\beta,{\rm ad}}~=~\la \hat{H}(0)\ra_\beta/b_{\rm ad}^2$, the nonadiabatic factor $Q^{\ast}(t)$ equals the ratio between the mean nonadiabatic energy and the corresponding adiabatic value. Therefore, values of $Q^{\ast}(t)>1$ indicate deviations from adiabatic dynamics and can be associated with quantum friction \cite{FK06}, which vanishes whenever $Q^{\ast}(t)~=~1$.

For the sake of illustration, we shall consider the harmonic Calogero--Sutherland model (CSM), a specific instance of Hamiltonian (\ref{Hsu11}). The CSM describes bosons confined in a time-dependent harmonic trap with inverse-square two-body interactions \cite{Calogero71,Sutherland71}, 
\beqa
 \hat{H}_{\rm CS}=\sum_{i=1}^{\N}\left[-\frac{\hbar^2}{2m}\frac{\partial^2}{\partial z_i^2}+\frac{1}{2}m\omega(t)^2 z_{i}^{2}\right]+\frac{\hbar^2}{m}\sum_{i<j}\frac{\lambda(\lambda-1)}{(z_i-z_j)^2},
 \label{eq:csm}
\eeqa
where $\lambda\geq0$ is the strength of the two-body interactions. 
While the many-body eigenstates are always symmetric under permutation of the particle coordinates and Pauli's exchange statistics is bosonic for all values of $\lambda$, 
the CSM represents an ideal gas of non-interacting particles obeying generalized (or fractional) exclusion statistics, a concept introduced by Haldane \cite{Haldane91} and further extended by Wu~\cite{Wu94}. 
In particular, for $\lambda=0$, the CSM reduces to the trapped ideal Bose gas. Similarly, for $\lambda=1$, the CSM is equivalent to a one-dimensional gas of bosons in the Tonks--Girardeau regime \cite{Girardeau60,GWT01} that obeys Fermi exclusion statistics as a result of the hard-core interactions. Therefore, when $\lambda=1$, the occupation number of a single particle state (e.g., of the trap) is restricted to the values $\{0,1\}$, and the 
 thermodynamics in this regime is identical to that of non-interacting spin-polarized trapped fermions.
As an upshot, the CSM allows one to account for the thermodynamics of non-interacting bosons and fermions on the same footing, given that it is governed by the exclusion statistics as opposed to the exchange statistics. 
More generally, for values of $\lambda$ other than zero and one, the CSM describes an ideal gas of Haldane anyons or ``geons'', particles with fractional exclusion statistics \cite{MS94}.
In the canonical ensemble, an explicit computation of the equilibrium mean-energy of a thermal state of the CSM~yields:
\begin{eqnarray}
\la \hat{H}(0)\ra_\beta=E_0(\N,\lambda) +\hbar\omega\sum_{k=1}^{\N} \frac{k}{e^{\beta k\hbar\omega}-1},
\label{eq:mean-csm}
\end{eqnarray}
where $E_0(\N,\lambda)$ denotes the ground-state energy: 
\beqa
E_0(\N,\lambda) = \frac{\hbar\omega}{2} \N[1 +\lambda(\N-1)].
\eeqa

Hence, the effect of the interactions on the thermodynamics is completely absorbed in the renormalization of the ground-state energy of the system \cite{Kawakami93}.

\subsection{Quantum Otto Cycle and Fundamental Limits}

\begin{figure}[h]
\begin{center}
\includegraphics[width=0.8\linewidth]{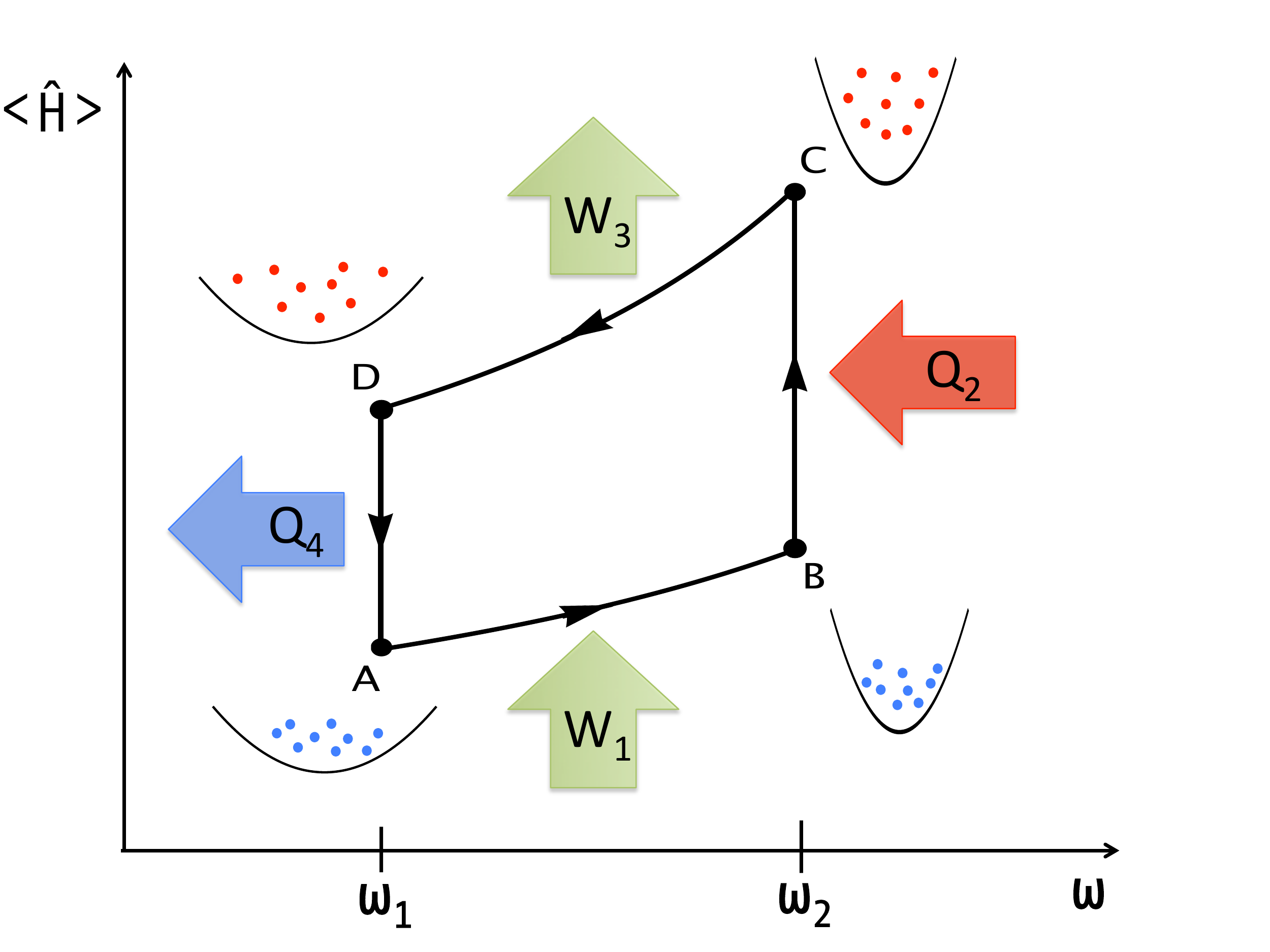}
\end{center}
\caption{{Many-particle quantum heat engine.} Quantum Otto cycle with a quantum fluid as a working medium confined in a harmonic trap with frequency varying between $\om_1$ and $\om_2$. States at A and C are thermal, while those at B and D are generally nonequilibrium states. }
\label{figure1}
\end{figure}

In the Otto cycle, isentropic and isochoric strokes alternate with each other, as illustrated in Figure~\ref{figure1}. There are two isentropes and two isochores per cycle, where energy in the form of work ($W$) and heat ($Q$) is transferred from and into the heat engine. In order of alternation, these are: \textit{isentropic compression} ($W_1$), \textit{hot isochore} ($Q_2$), \textit{isentropic expansion} ($W_3$) and \textit{cold isochore} ($Q_4$). The isentropic strokes involve a change in the trapping frequency at constant temperature, while during an isochore, the temperature varies and the frequency kept constant. The efficiency of a heat engine is defined as the total output work per input heat:
\begin{eqnarray}\label{Efficiency}
\eta=-\frac{\langle W_1\rangle+\langle W_3\rangle}{\langle {\rm Q}_2\rangle},
\end{eqnarray}
where $\langle W_{1(3)}\rangle=\langle H\rangle_{B(D)}-\langle H\rangle_{A(C)}$ and $\langle {\rm Q}_{2(4)}\rangle=\langle H\rangle_{C(A)}-\langle H\rangle_{B(D)}$. The heat engine is decoupled from the thermal reservoir during the isentropic strokes, when the dynamics is unitary. We use the scaling dynamics of the working medium (Equation \eqref{eq:csm}) to predict the mean energy according to Equation \eqref{aven0}.
The efficiency of the many-particle quantum heat engine run in finite time is given by:
\begin{eqnarray}\label{Efficiency}
\eta=1-\frac{\omega_1}{\omega_2}\left(\frac{Q_{CD}^{\ast}\langle H\rangle_C-\frac{\omega_2}{\omega_1}\langle H\rangle_A}{\langle H\rangle_C-Q_{AB}^{\ast}\frac{\omega_2}{\omega_1}\langle H\rangle_A}\right),
\end{eqnarray}
where the mean energy of the thermal states $A$ and $C$ corresponds to $\langle H\rangle_A=\la H(0)\ra_{\beta_{\rm c}}$ with $\omega=\omega_1$ and $\langle H\rangle_C=\la H(0)\ra_{\beta_{\rm h}}$ with $\omega=\omega_2$; see Equation \eqref{eq:mean-csm}.

Equation (\ref{Efficiency}) reduces in the adiabatic limit ($Q^{\ast}_{AB(CD)}\rightarrow 1$) to the Otto efficiency: 
\beqa
\eta_O=1-\frac{\omega_1}{\omega_2}\ ,
\eeqa
which is shared as an upper bound by both single- and many-particle quantum Otto cycles. 

When a compression (expansion) of finite duration $\tau$ ($\tau'$) is considered, such that the frequency monotonically varies as a function of time, we have $Q^{\ast}_{AB(CD)}\leq Q^{\ast}_{\rm sq}$, and Equation \eqref{Efficiency} further implies that the finite-time efficiency $\eta$ is bounded from below and above as:
\begin{equation}\label{efficiencybound1}
\eta_{\text{sq}}\leq \eta \leq \eta_O,
\end{equation}
where $\eta_{\text{sq}}$ is the efficiency under a sudden quench of the trap frequency for which $Q^*_{\text{sq}}~=~(\omega_1^2~+~\omega_2^2)/(2\omega_1\omega_2)$ \cite{Husimi53}. This applies to driving protocols, such that
$\omega(0)=\omega_1$ and $\omega(\tau)=\omega_2$ in the compression stroke and with $\omega(0)=\omega_2$ and $\omega(\tau')=\omega_1$ in the expansion stroke. For a monotonous frequency, we have that $Q^*(\tau)\rightarrow Q^*_{\text{sq}}$ in the limit $\tau\rightarrow 0$, as the $\tau$-parameterization of a given protocol $\omega(t),\ \tau\in[0,+\infty]$ approaches then a smooth representation of the step function (e.g., $\omega_1\theta(-t)+\omega_2\theta(t)$ for the compression). 
In addition, as proven in \cite{JBdC15}, the~efficiency (Equation~\eqref{Efficiency}) is bounded by a non-adiabatic Otto limit,
\begin{equation}\label{efficiencybound2}
\eta \leq \eta_{\textrm{nad},O}\equiv 1-Q_{CD}^{\ast}\frac{\omega_1}{\omega_2}\ ,
\end{equation}
independent of the number of particles $\N$ and interaction potential $V$.

This formula encodes the ``tragedy of finite-time thermodynamics'' in the many-particle setting: 
the maximum efficiency is achieved under slow driving, in the adiabatic limit,
when the QHE operates at 
vanishing output power $-(\la W_1\ra+\la W_3\ra)/\tau_c$ as a result of the requirement for a long cycle time $\tau_c$. 
By contrast, realistic engines operated in finite time achieve a finite output power at the cost of introducing 
nonadiabatic energy excitations that represent quantum friction and lower the efficiency. That nonadiabatic effects generally decrease the engine efficiency follows from the fact that $Q_{AB(CD)}^{\ast}\geq 1$.
However, we will show that this tradeoff is not fundamental in nature and can be~avoided.

\subsection{Optimization of Power and Efficiency}\label{Opt}

\subsubsection{General Method}\label{GeneralMeth}

We first provide a general method to optimize the power of the QHE for any finite-time protocol, which will prove useful in deriving both analytical and numerical results. 
As shorthand, we introduce the ratio of trap frequencies:
\beqa 
x \equiv \frac{\omega_1}{\omega_2} \leq1,
\eeqa
as well as the ratio of temperatures of the cold and hot reservoir:
\beqa
a \equiv \frac{\beta_{\rm h}}{\beta_{\rm c}} \leq1.
\eeqa

For a fixed cycle time $\tau_c$, we optimize the output power $P=|W|/\tau_c$, where the total output work $|W| = -(W_1+W_3) $ is given by:
\begin{equation}\label{WorkTD}
|W| = (1-x^{-1}Q_{AB}^{\ast})\langle H\rangle_A + (1-x Q_{CD}^{\ast})\langle H\rangle_C\ .
\end{equation}

Assuming the hot bath to be at high temperature, \emph{i.e.}, $\sigma_{\rm h}\equiv \N\hbar\beta_{\rm h}\omega_2\ll 1$, the optimal work is reached when the frequency ratio $x$ is a solution of the following equation:
\begin{equation}\label{xoptTDM}
\frac{a\beta_{\rm c} \langle H\rangle_A}{N}=N_F(x)\ x^2,
\end{equation}
with the factor $N_F(x)$, dependent on $\tau$ and $\tau'$, given by:
\begin{equation}\label{NF}
N_F(x)=\frac{Q_{CD}^{\ast}+x(\partial_x Q_{CD}^{\ast})}{Q_{AB}^{\ast}-x(\partial_x Q_{AB}^{\ast})}\ .
\end{equation}

We note that for an adiabatic protocol, $N_F=1$, while for a sudden quench driving, we have $N_F=x^2$. 
Further, for an adiabatic protocol in the high temperature regime, the mean energy $\langle H\rangle_A$ in Equation~\eqref{xoptTDM} is shifted by the effective zero-point energy term $-E_0(\N,\lambda)+E_0(\N,1/2)$, which results from the expansion of $\langle H\rangle_C$ in Equation \eqref{WorkTD} at high temperature. 

In \cite{JBdC15}, it has been shown that for the sudden quench regime, the efficiency at optimal power can be drastically enhanced for a large number of particles, when $\sigma_{\rm c(h)}\equiv \N\hbar\beta_{\rm c(h)}\omega_{1(2)}> 1$ is achieved at relatively low temperatures, keeping $\sigma_{\rm c(h)}\ll \N$. 
In this regime, one can write an explicit approximation for the total work per cycle:
\begin{equation}\label{WlargeN}
|W|=\frac{\N}{\beta_{\rm c}}\left[\left(1-x^{-1}Q_{AB}^{\ast}\right)\mu_\lambda(\sigma_{\rm c})+\left(1-xQ_{CD}^{\ast}\right)\frac{1}{a}\mu_\lambda\left(\frac{a}{x}\sigma_{\rm c}\right)\right],
\end{equation}
where we estimated the Riemann sum in \eqref{eq:mean-csm} by a Riemann integral using $\hbar\beta\omega\ll1$ (as the corrections are bounded by $\N(\hbar\beta\omega)^2/2$ and $\N$ cannot be too large), with:
\begin{equation}\label{mu}
\mu_\lambda(\sigma)=\frac{1}{\sigma}\int_{0}^{\sigma}ds\frac{s}{e^s-1}+\sigma\frac{\lambda}{2}
=\frac{\pi^2}{6\sigma}+\log{(1-e^{-\sigma})}-\sigma^{-1}\text{Li}_2(e^{-\sigma})+\frac{\lambda}{2}\sigma\ .
\end{equation}

Here, we use the definition of the standard polylogarithm functions \mbox{$\text{Li}_n(z)=\sum_{j=1}^{\infty}j^{-n}z^j,$}\mbox{$\ n\geq 1$~\cite{Abramowitz}. }
The function $\mu_\lambda(\sigma)$ monotonically decreases for $\lambda=0$, while it monotonically increases for all $\lambda\geq 1/2$. For $0<\lambda<1/2$, $\mu_\lambda(\sigma)$ decreases until reaching a minimum at $\sigma=\sigma_0(\lambda)$ and increases monotonically for larger values of $\sigma$. One can interpret the function $\mu_\lambda(\sigma_{\rm c(h)})$ as the quantum deviation of the mean energy $\langle H\rangle_{\rm A(C)}$ from its classical value $\N/\beta_{\rm c(h)}$. For any given protocol, Equation \eqref{WlargeN} allows one to compute the optimal frequency ratio $x$ that maximizes the work done per cycle. 

\subsubsection{Optimizing Adiabatic Output Power}\label{optadiab}

In this section, we assume that the adiabaticity condition is fulfilled for a fixed value of the duration of the strokes $\tau$. Optimizing the output power is then equivalent to maximizing the output work. We shall provide expressions for the maximum adiabatic output work $\tilde{W}_{\rm ad}^{(\N)}$ and the adiabatic efficiency at optimal output power $\eta_{\rm ad}^{(\N)}$ of the scaled-up QHE with $\N$ particles as the working medium. Our results provide a generalization of the optimization of the single-particle \mbox{adiabatic QHE \cite{Abah12,CA75}. }

Given that for an adiabatic driving, the nonadiabatic factor reduces to unity $Q^*=1$, the total output work denoted by $|W_{\text{ad}}|$ becomes independent of $\lambda$ and depends only on the thermal energy, according to Equation \eqref{WorkTD}. 
 
Assuming $\sigma_{\rm h}=\N\beta_{\rm h}\hbar\omega_2\ll 1$, we obtain:
\beqa
|W_{\text{ad}}|=\left(1-\frac{1}{x}\right)E_{\N,1/2}(\omega_1,\beta_{\rm c})+(1-x)\frac{\N}{a\beta_{\rm c}}\ ,
\eeqa
where the function $E_{\N,\lambda}(\omega_1,\beta_{\rm c})$ is given by \eqref{eq:mean-csm} for any $\lambda\geq 0$. To obtain the previous equation, we use a Taylor expansion of the energy $\langle H\rangle_C$ for small $\sigma_{\rm h}$. The optimal frequency ratio is found to be $x\approx x_{\rm opt}^{(\N)}\equiv\left(a\beta_{\rm c} E_{\N,1/2}(\omega_1,\beta_{\rm c})/\N\right)^{1/2}$, a function of the temperature ratio $a=\beta_{\rm h}/\beta_{\rm c}$. The maximum work and the efficiency at optimal output power read:\begin{subequations}\label{WoptAd}
\begin{eqnarray}
\tilde{W}_{\text{ad}}^{(\N)}&=& \frac{\N}{a\beta_{\rm c}}\left(1-x_{\rm opt}^{(\N)}\right)^2\ ,\label{WoptAd1}\\
\eta_{\text{ad}}^{(\N)}&=& 1-x_{\rm opt}^{(\N)}\ .
\label{WoptAd2}
\end{eqnarray}
\end{subequations}

Using the asymptotic expansion of the thermal energy, one can derive the asymptotic expressions for the optimal work and the corresponding efficiency in different regimes distinguished by the value of $\sigma_{\rm c}$.
Taking the cold reservoir at high temperature $\sigma_{\rm c}\ll 1$, it is found that $E_{\N,1/2}(\omega_1,\beta_{\rm c})~\approx~\frac{\N}{\beta_{\rm c}}\left(1+O(\sigma_{\rm c}^2)\right)$ and $x_{\rm opt}^{(\N)}\approx a\left(1~+~O(\sigma_{\rm c}^2)\right)$. The engine exhibits then a classical performance characterized by:
\begin{subequations}\label{OptWorkEffAdHT}
\begin{eqnarray}
\tilde{W}_{\text{ad}}^{(\N)}&\approx &\frac{\N}{a\beta_{\rm c}}\left\{\left(1-\sqrt{a}\right)^2+O(\sigma_{\rm c}^2)\right\}\ ,\label{OptWorkEffAdHT1}\\
\eta_{\text{ad}}^{(\N)}&\approx & 1-\sqrt{a}+O(\sigma_{\rm c}^2)\ .\label{OptWorkEffAdHT2}
\end{eqnarray}
\end{subequations}

The latter equations show that many-particle effects are negligible for an adiabatic protocol when the number of particles is not too large, so that $\sigma_{\rm c(h)}\ll 1$. The efficiency is then set by the Curzon--Ahlborn result, $1-\sqrt{a}$, associated with a classical heat engine operated in the \mbox{adiabatic limit \cite{CA75}. }

It has been shown recently that increasing the particle number can lower efficiency drastically; see \cite{JBdC15}. Indeed, assuming that $\sigma_{\rm h}\ll 1$ and $1\leq\sigma_{\rm c}\ll \N$, by Equations \eqref{WlargeN} and \eqref{mu}, we find that: 
\begin{subequations}\label{OptWorkEffAdInt}
\begin{eqnarray}
\tilde{W}_{\text{ad}}^{(\N)} &\approx & \frac{\N}{a\beta_{\rm c}}\left(1-\sqrt{a\ \mu_{1/2}(\sigma_{\rm c})}\right)^2\label{OptWorkEffAdInt1},\\ 
\eta_{\text{ad}}^{(\N)} &\approx & 1-\sqrt{a\ \mu_{1/2}(\sigma_{\rm c})} \ ,\label{OptWorkEffAdInt2}
\end{eqnarray}
\end{subequations}
where $\mu_{1/2}(\sigma_{\rm c})$ is greater than one and increases monotonically with $\sigma_{\rm c}$. Consequently, the efficiency is lesser than the Curzon--Ahlborn efficiency, $1-\sqrt{a}$. Equations \eqref{OptWorkEffAdInt1} and \eqref{OptWorkEffAdInt2} are valid only for small values of $a=\beta_{\rm h}/\beta_{\rm c}$ consistent with $\sigma_{\rm h}\ll1$. This many-particle applies for a large number of particles $\N\gg (\hbar\beta_{\rm c}\omega_1)^{-1}$, keeping the temperature of the cold reservoir relatively small $\hbar\omega_1\beta_{\rm c}\ll 1$. 

Now, considering the cold reservoir at very low temperature $\sigma_{\rm c}\gg \N$ (\emph{i.e.}, $\hbar\omega_1\beta_{\rm c}\gg 1$), the thermal energy is negligible, and we find $x_{\rm opt}^{(\N)}\approx \sqrt{\frac{\hbar\omega_1\beta_{\rm h} (\N+1))}{4}}$, leading to:
\begin{subequations}\label{OptWorkEffAdVLT}
\begin{eqnarray}
\tilde{W}^{(\N)}_{\text{ad}}&\approx & \frac{\N}{a\beta_{\rm c}}\left(1-\sqrt{\frac{\hbar\omega_1\beta_{\rm h} (\N+1)}{4}}\right)^2\ ,\label{OptWorkEffAdVLT1}\\
\eta_{\text{ad}}^{(\N)}&\approx & 1-\sqrt{\frac{\hbar\omega_1\beta_{\rm h} (\N+1)}{4}}\ .\label{OptWorkEffAdVLT2}
\end{eqnarray}
\end{subequations} 

In this quantum limit, the performance of the engine decreases drastically compared to the classical limit, as the term in the square root of Equations \eqref{OptWorkEffAdVLT1} and \eqref{OptWorkEffAdVLT2} is much larger than the corresponding term $a=\beta_{\rm h}/\beta_{\rm c}$ in Equations \eqref{OptWorkEffAdHT1} and \eqref{OptWorkEffAdHT2}.

From the comparison of these limiting cases, it is found that in the adiabatic limit, the efficiency and total work output are maximized in the classical regime when both the cold and hot reservoirs are at high temperature, e.g., $\sigma_{\rm c(h)} \ll1$. It is in this regime that it is most convenient to operate the scaled-up QHE, avoiding quantum corrections that lower the performance.
The work per cycle is then scaled by a factor of $\N$ with respect to the single-particle QHE, and the efficiency is then set by the Curzon--Ahlborn efficiency, according to Equations (\ref{OptWorkEffAdHT1}) and (\ref{OptWorkEffAdHT2}). Next, we discuss how to achieve this adiabatic performance under nonadiabatic STA protocols that shorten the cycle \mbox{operation time.}

\section{Superadiabatic Many-Particle Quantum Heat Engines}

As discussed in the Introduction, it has recently been shown that it is possible to boost the performance of a single-particle QHE by reducing the cycle time, replacing the compression and expansion strokes of a quantum Otto cycle by STA protocols \cite{Zhang14,delcampo14}. 
 In what follows, we shall explore the optimization of the finite-time thermodynamics of a many-particle QHE using the STA for many-body systems \cite{delcampo11b,Choi11,Schaff11,DB12,DRZ12,delcampo13,DJD14,Saberi14,Rohringer15} 
 to engineer a scale-up many-particle QHE operating at maximum efficiency and with high output power. 
 We shall introduce two different approaches depending on the given specifications of the many-particle QHE.

\subsection{First Approach: Finite-Time Optimization and Accidental STA}\label{AccSTA}

Consider the operation of the many-particle QHE with given cold and hot thermal reservoirs at fixed (inverse) temperatures $\beta_{\rm c}$ and $\beta_{\rm h}$.
Assuming that a specific time-dependent protocol $\om(t)$ describes the evolution of the working medium, the conditions for optimal performance can be determined by maximizing the output power for a variable and finite duration $\tau$ of the compression and expansion strokes. The duration of the isochores constitutes an overhead that adds to the total cycle time $\tau_c$ and plays no role in our analysis.
The possibility to engineer a superadiabatic many-particle QHE with zero friction
relies on the existence of protocols $\om(t)$ during which the nonadiabatic dynamics become adiabatic at a specific set of values for the duration of the strokes $\{\tau_n,\ n= 1,2,3,\dots\}$, which depends on the frequencies $\omega_{1},\ \omega_2$ of the trap. We refer to any such instance as an \textit{accidental STA}. 

To illustrate this approach, we first introduce an accidental STA protocol that interpolates between the adiabatic and the sudden quench limit ($\tau=0$). 
In particular, we choose a modulation of the trap frequency for which the adiabaticity coefficient $\dot{\om}/\om^2$ \cite{LR69} remains constant during the compression and expansion strokes. We mention that this model has been introduced in the context of quantum refrigerators \cite{Rezek09}, noisy QHE \cite{Kosloff13,Stefanatos14} and the dynamics of a single particle in a trap \cite{Kosloff14}. Here, we introduce this modulation of the trap frequency following a different approach that suits better the many-particle QHE model Equation \eqref{eq:csm}. The condition $|\dot{\omega}(t)/\omega(t)^2|=2\gamma^{-1}$ with $\gamma>0$ is satisfied by
the time-dependent frequencies:
\begin{subequations}\label{accprot}
\begin{eqnarray}
\omega_{AB}(t)=\om_1 t_1/(t_1-t),\ t\in [0,\tau]\label{accprotAB}
,\\
\omega_{CD}(t)=\om_2 t_2/(t+t_2),\ t\in [0,\tau]\label{accprotCD},
\end{eqnarray}
\end{subequations}
where $\om_1,\om_2,t_1,t_2$ and $\tau$ are fixed by the boundary conditions
$\omega_{AB}(0)~=~\om_1,\ \omega_{AB}(\tau)~=~\om_2$ and~$\omega_{CD}(0)~=~\om_2,\ \omega_{CD}(\tau)~=~\om_1$, leading to $t_1~=~(1-\omega_1/\omega_2)^{-1}\tau$ and~$t_2~=~(\omega_1/\omega_2)t_1$. Notice that the protocol is symmetric $\omega_{CD}(t)=\omega_{AB}(\tau-t)$ and can be rewritten as $\omega_{CD}(t)~=~\omega_1\omega_2\tau/(\omega_1\tau~+~t(\omega_2-\omega_1))$. The dimensionless parameter $\gamma$ is given by $\gamma=2\om_1t_1$ and can be directly related to the duration of the stroke:
\begin{equation}\label{taugamma}
\gamma =\frac{2\om_1\tau}{1-x},
\end{equation}
via the frequency ratio $x\equiv \omega_1/\omega_2$.

\begin{figure}[h]
\begin{center}
\includegraphics[width=0.4\linewidth]{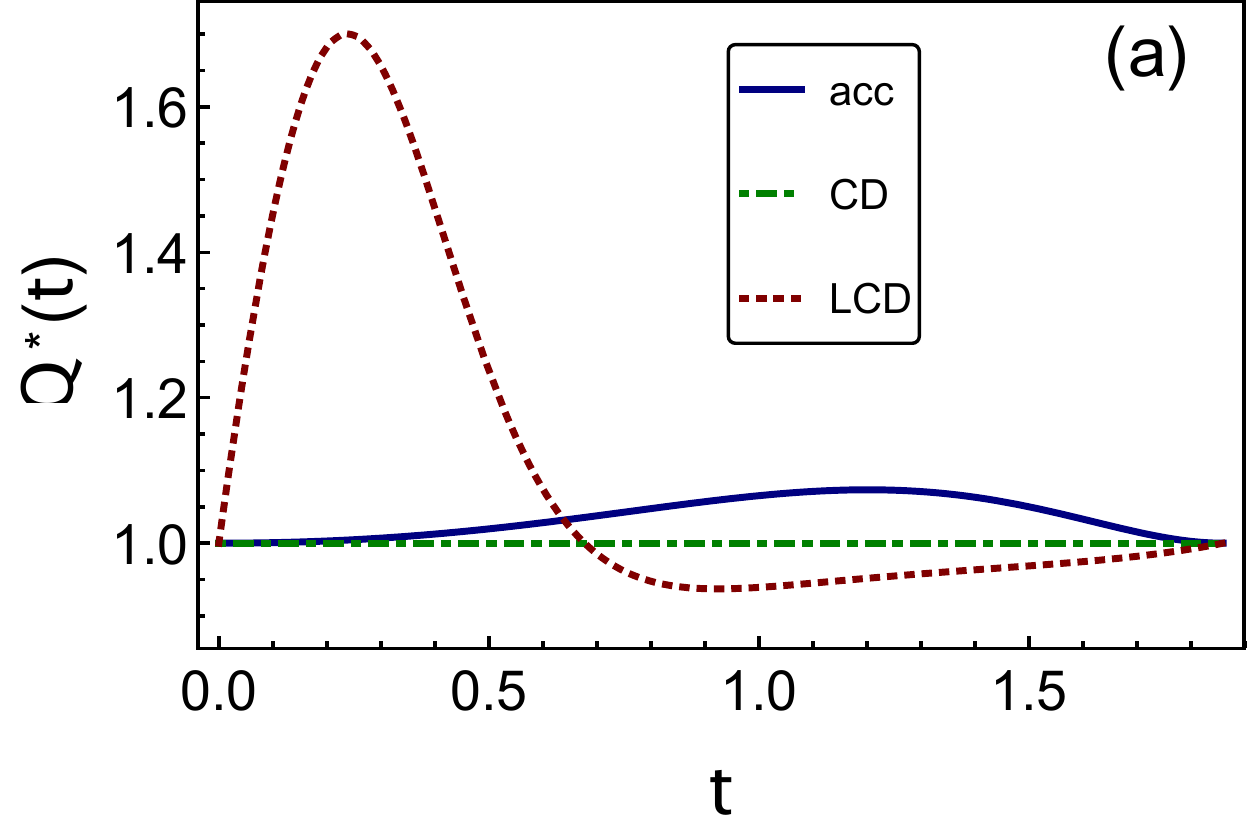}\includegraphics[width=0.4\linewidth]{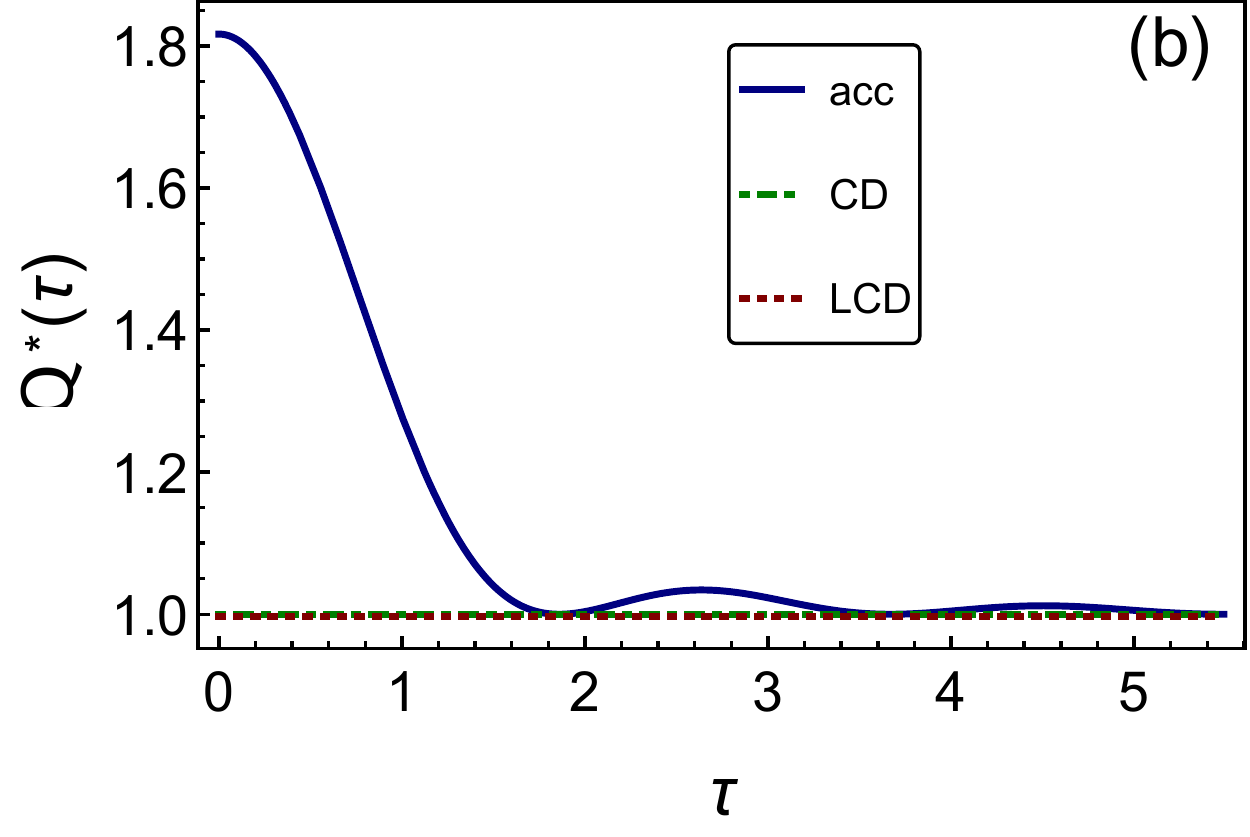}
\end{center}
\caption{{Nonadiabatic coefficient during a shortcut to adiabaticity.} (\textbf{a}) Nonadiabatic coefficient during an expansion/compression engineered via accidental shortcut to adiabaticity (STA) (acc), counterdiabatic driving (CD; discussed in Section \ref{secCD}) and local counterdiabatic driving (LCD; see \mbox{Section \ref{secLCD})} as a function of the evolution time $0\leq t/\tau_1\leq 1$, where $\tau_1$ is the minimum duration for an accidental STA, given in Equation \eqref{taun} (take $n=1$). (\textbf{b}) Variation of the nonadiabatic factor as a function of the duration $\tau$ of the protocols, in units of $\omega_1^{-1}$. In ({a}) and ({b}), we take $\omega_1/\omega_2=0.3$. }
\label{figureQ}
\end{figure}

\begin{figure}[h]
\begin{center}
\includegraphics[width=0.33\linewidth]{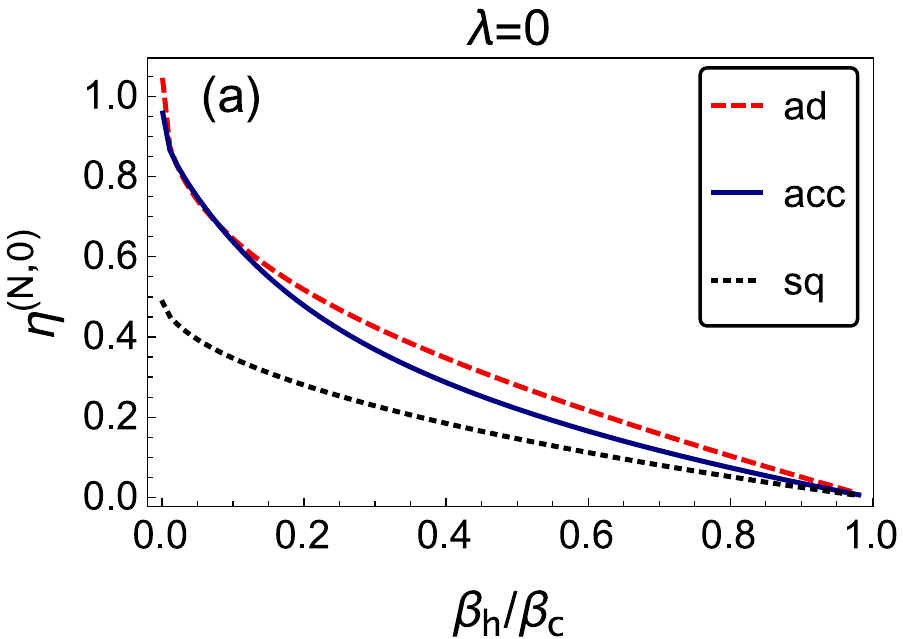}\includegraphics[width=0.33\linewidth]{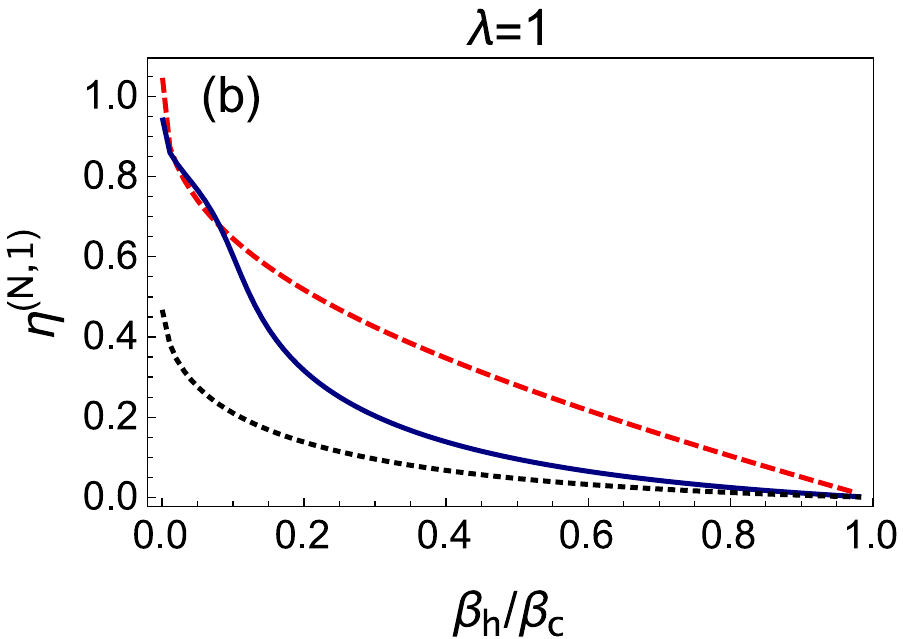}\includegraphics[width=0.33\linewidth]{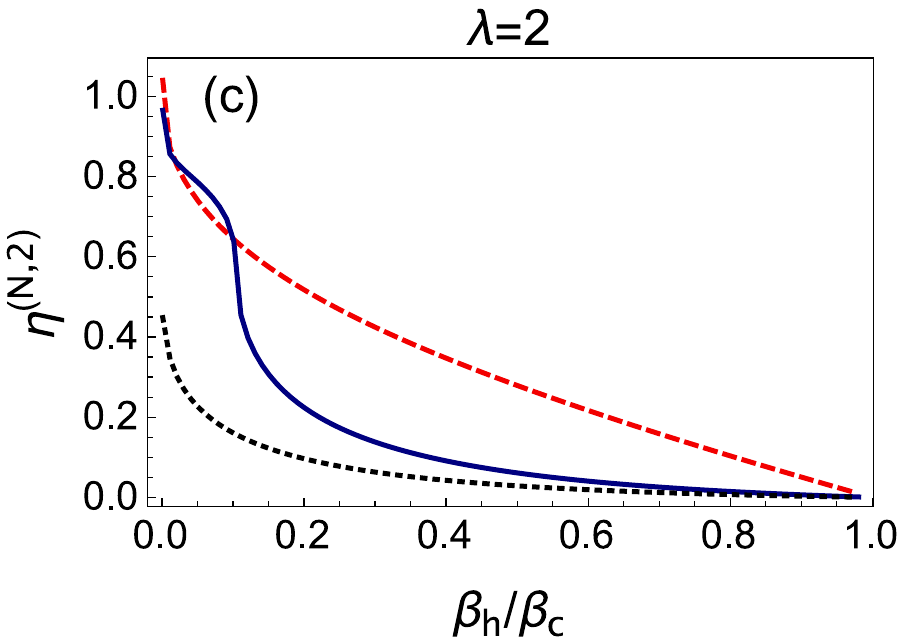}
\linebreak\\
\includegraphics[width=0.33\linewidth]{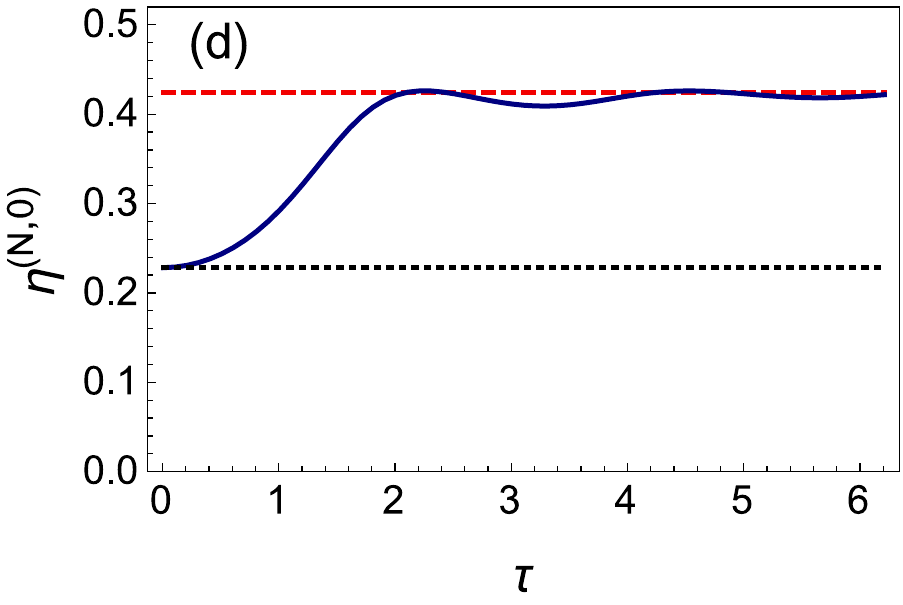}\ \includegraphics[width=0.32\linewidth]{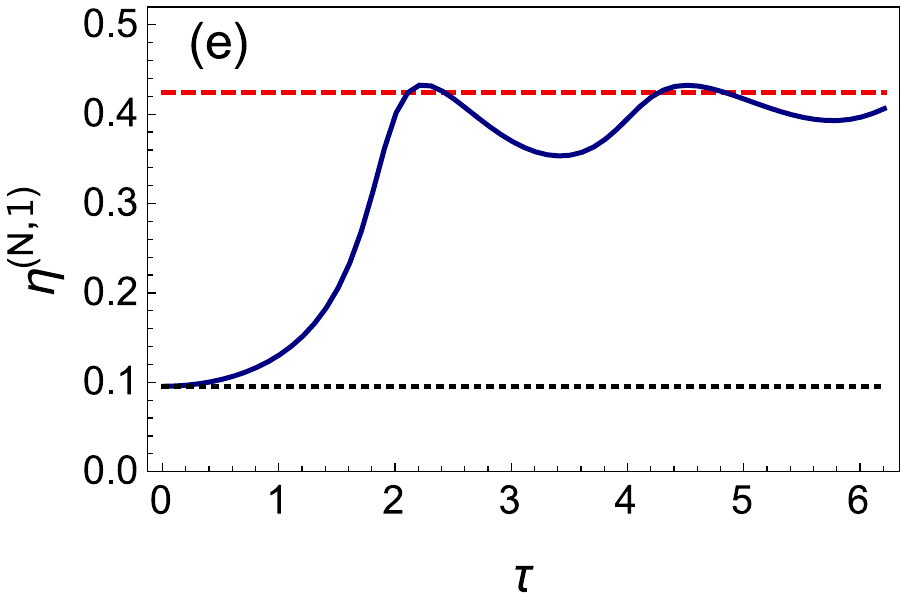}\ \includegraphics[width=0.32\linewidth]{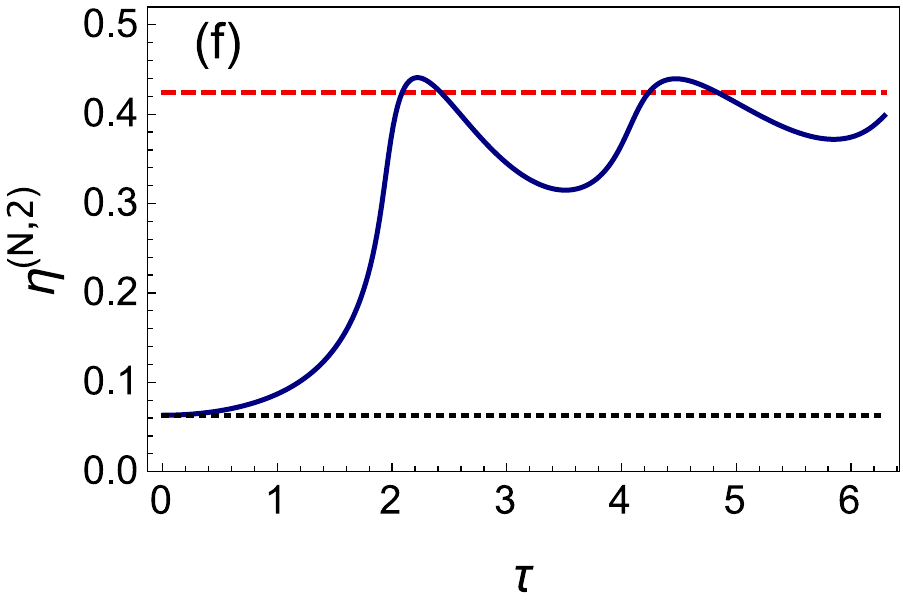}
\end{center}
\caption{{Efficiency of a many-particle quantum heat engine (QHE) run in finite-time at optimal power.} Top panels (\textbf{a}--\textbf{c}) display the efficiency of a many-particle QHE at maximum power as a function of $\beta_{\rm h}/\beta_{\rm c}$ for the accidental protocol (acc) with $\tau=1.5/\omega_1$, a sudden quench (sq) and adiabatic driving (ad). The interaction strength takes values $\lambda=0,1,2$ from left to right and $\N=500$. 
 A similar representation as a function of $\tau$, in units of $\omega_1^{-1}$, is done in (\textbf{d}--\textbf{f}), where $\beta_{\rm h}/\beta_{\rm c}=0.3$.
 In both representations, a transition is observed from the sudden-quench to the adiabatic limits that is governed by the nonadiabatic coefficient computed via Equation \eqref{QTD}. 
}
\label{figure3}
\end{figure}

The accidental STA with finite $\gamma$ interpolates between the sudden quench ($\gamma\rightarrow 0$) and adiabatic limit ($\gamma\rightarrow +\infty$). 
This is verified by exact computation of the nonadiabatic factors $Q^*_{AB}$ and $Q^*_{CD}$ that happen to be equal and given by (see Figure \ref{figureQ}):
\begin{equation}\label{QTD}
Q^{*}=\left\{
\begin{aligned}
 & 1+\frac{\cosh{(\sqrt{1-\gamma^2}\ln(x))}-1}{1-\gamma^2},\ \text{if } \gamma\leq 1,\\
 & 1+\frac{1}{2}\ln(x)^2\ \ \ \ \ \ \ \ \ \ \ \ \ \ \ \ \ \ \ \ \ \ \ \ \ \ \ ,\ \text{if } \gamma= 1,\\
 & 1+\frac{1-\cos{(\sqrt{\gamma^2-1}\ln{(x)})}}{\gamma^2-1}\ ,\ \text{if } \gamma\geq 1,
\end{aligned}
\right.
\end{equation}
where $x=\omega_1/\omega_2$. In Figure \ref{figureQ}a, we plot the nonadiabatic factor Equation \eqref{QTD} as a function of the time of evolution $t$ for a given duration $\tau$ of the compression/expansion stroke and as a function of $\tau$ in Figure \ref{figureQ}b. 
For the optimization of the QHE, the value of $x$ is kept fixed, while $\tau$ is varied.
When $\tau\rightarrow 0$, we find that $Q^*\rightarrow (x^2+1)/(2x)$, equal to the sudden quench factor $Q^*_{\rm sq}$.
On the contrary, we obtain the adiabatic factor $Q_{\text{ad}}^{\ast} = 1$ in the limit $\tau\rightarrow +\infty$. 
By Equation \eqref{QTD}, for $\gamma\leq 1$ (or equivalently $\tau\leq (1-x)/(2\omega_1)$), the nonadiabatic factor decreases monotonically from the sudden quench value to a lower one that is still above unity. In Figure \ref{figure3}, one can see that the efficiency behaves as in a sudden quench protocol; see also Figure \ref{figureQ}b to see the relation with the nonadiabatic factor. Yet, for $\gamma>1$ (or equivalently $\tau > (1-x)/(2\omega_1)$), the nonadiabatic factor oscillates above the adiabatic value; see Figure \ref{figureQ}b. At the contact points (obtained in \cite{Rezek09,Kosloff14} by a different method), $\gamma_n=(1+4\pi^2 n^2/\ln{(x)}^2)^{1/2}$, with $ n=1,2,3\dots$, we find an STA given that nonadiabatic factor equals unity, $Q^{*}=1$. At these points, the work and the efficiency reproduce the corresponding values found in the slow driving limit. For a given value of $x$, the accidental STA times $\tau_n$ are found to be: 
\begin{equation}\label{taun}
\tau_n=\frac{1-x}{2\omega_1}\sqrt{1+\frac{4\pi^2 n^2}{\ln{(x)}^2}}.
\end{equation} 

Using Equation (\ref{taun}) to determine the duration of the compression and adiabatic strokes leads to a many-particle QHE with zero friction, maximum efficiency and reduced cycle time with respect to the adiabatic case.
Two fundamental bounds are found on the minimum time $\tau_1$ for the existence of an STA under this driving protocol:
\begin{equation}\label{tau1bounds}
\frac{1}{2\omega_1}\leq\tau_1\leq\frac{\pi }{\omega_1}\ ,
\end{equation}
where we used that $(1-x)^{-2}\leq 1+(4\pi^2\ln(x)^{-2})\leq 4\pi^2(1-x)^{-2}$. Notice that Equation \eqref{tau1bounds} is independent of $\N$ and $\lambda$. These bounds are sharp for $x\sim 1$, $\tau_n\sim \frac{\pi n}{\omega_1}$ and for $x\sim 0$, $\tau_n\sim 1/(2\omega_1)$. In Appendix A, it is shown that: 
\beqa
b(\tau_n)=\sqrt{x}=\sqrt{\frac{\omega_1}{\omega_2}}=b_{\rm ad},\ \dot{b}(\tau_n)=\ddot{b}(\tau_n)=0,
\eeqa
as required for an STA protocol.

Concerning many-particle effects, we point out that the efficiency at optimal power can be enhanced drastically for $\tau\leq 1/\omega_1$ if $\lambda=0$ (a similar conclusion holds for $\lambda\ll 1$, typically $\lambda<0.1$) where $1/\omega_1$ characterizes the transition between the sudden quench and adiabatic regime. Indeed, taking $\tau>1/\omega_1$, the nonadiabatic many-particle efficiency at optimal power becomes closer to the adiabatic efficiency at optimal power; see Figure \ref{figure3}. 

Surprisingly, the nonadiabatic efficiency at optimal output power can beat the adiabatic value for $\tau_1'<\tau<\tau_1$, where $\tau_1'$ and $\tau_1$ refer respectively to the first and the second intersection between the nonadiabatic and adiabatic efficiency; see Figure \ref{figure3}. This effective enhancement can be explained by the fact that the optimal value of $\omega_1/\omega_2$ is smaller in the nonadiabatic case, for $\tau_1'<\tau<\tau_1$. 
However, the nonadiabatic efficiency is still bounded by the Otto limit, \emph{i.e.}, the adiabatic efficiency evaluated for the same frequency ratio. As already mentioned, in Figure \ref{figure3}, we find that for $\tau=\tau_1$, as well as for $\tau=\tau_1'$, the efficiency at output power of the nonadiabatic protocol equals the value of an adiabatic protocol. For this to happen, $\tau_1$ and $\tau_1'$ do not need to correspond to a proper STA with a nonadiabatic factor equal to one. Indeed, $\tau_1'$ (first intersection) does not. To find out the value of the minimum adiabatic shortcut time, one has to find the optimal frequency ratio for an adiabatic protocol (following the approach described in Section \ref{optadiab}) and substitute this value into Equation~\eqref{taun} taking $n=1$. We have verified numerically that $\tau_1$ is the first accidental STA. For $\tau=\tau_1'$, the nonadiabatic efficiency of the scaled-up QHE matches the adiabatic value; yet, $Q^{\ast}>1$.

\subsection{Second Approach: Adiabatic Optimization and STA}\label{secAD}

In this section, we introduce an alternative approach to boost the performance of a many-particle QHE via STA. It exploits the possibility of engineering 
an STA that mimics an adiabatic expansion and compression stroke in an arbitrary fixed amount of time $\tau$, between any two values of the frequency of the trap, $\om_1$ and $\om_2$. This allows one to opt first for an optimization of the QHE parameters assuming the dynamics to be adiabatic. In particular, for given temperatures of the hot and cold reservoirs, the frequency ratio is chosen to maximize the adiabatic output power. This boils down to maximizing the total work per cycle, as the time of the compression and expansion strokes is fixed; see Section~\ref{optadiab}. The assumption of adiabatic dynamics is then effectively fulfilled by replacing the adiabats by STA protocols. This second approach is more versatile as the duration of the unitary strokes is not restricted to specific values and can be pre-scheduled.
We shall consider three different techniques to engineer the required STA.

\subsubsection{Reverse Engineering of the Scaling Dynamics}\label{secRI}

Via reverse engineering, one can formulate a general theory of STA thanks to the consistent description of the dynamics given by Equations \eqref{Ermakoveq} and \eqref{Qstar}. Indeed, if one assume that there exists a time $\tau$, such that the scaling factor $b(t)$ satisfies the boundary conditions \cite{Chen10,delcampo11b}:
\begin{subequations}
\begin{eqnarray}
&&b(0)=1,\ \dot{b}(0)=0,\ \ddot{b}(0)=0\label{bcondt0},\\ 
&&b(\tau)=b_{\rm ad},\ \dot{b}(\tau)=0,\ \ddot{b}(\tau)=0\label{bcondtau},
\end{eqnarray}
\end{subequations}
 it follows that: 
\beqa
Q^{*}(\tau)=1. 
\eeqa

In particular, this is consistent with the boundary conditions that $\omega(t)$ must satisfy in the Ermakov Equation \eqref{Ermakoveq}, which we rewrite as:
\begin{equation}\label{Ermarkov2}
\omega(t)^2=\frac{\omega_0^2}{b(t)^4}-\frac{\ddot{b}(t)}{b(t)}\ ,
\end{equation}
and where $\omega(t)$ is the solution of the equation above with $\omega(0)=\omega_0$ and $\omega(\tau) = \omega_0/b_{\rm ad}^2$.

The conditions given by Equations \eqref{bcondt0} and \eqref{bcondtau} are satisfied if the scaling factor takes the following form:
\begin{equation}\label{bpolyn}
b(t)=1+a_3t^3+a_4t^4+a_5t^5\ ,
\end{equation}
where $a_3=10(b_{\rm ad}-1)/\tau^3$, $a_4=-15(b_{\rm ad}-1)/\tau^4$, and $a_5=6(b_{\rm ad}-1)/\tau^5$; see \cite{Chen10,delcampo11b}. However, from Equation \eqref{Ermarkov2}, it is not clear whether $\omega(t)$ is real for all $t\in[0,\tau]$. There must be a condition on $\tau$ (for $\omega_0$ and $b_{\rm ad}$ fixed) to ensure the positivity of $\omega(t)^2$. It is sufficient to require $\ddot{b}(t)b(t)^3~\leq~\omega_0^2,\ \forall t~\in~[0,\tau]$, which leads to a non-trivial polynomial equation. A numerical analysis shows that there exists a unique solution $\tau_c$, such that $\forall\tau>\tau_c$ and $\forall t\in[0,\tau]$; we have $\omega(t)^2\geq 0$. 
This~approach generalizes the one pursued to boost the performance of a single-particle QHE in \cite{delcampo14}. 

Equation \eqref{bpolyn} is one instance of STA constructed via reverse engineering of the scaling dynamics. The protocol defined by Equations \eqref{accprotAB} and \eqref{accprotCD} also satisfies the condition \mbox{Equations \eqref{bcondt0} and \eqref{bcondtau};} see Appendix A. Following this observation, 
one can construct many other driving protocols leading to accidental STA, such as:
\begin{equation}\label{bSTAex}
b(t)=1+\beta\left\{\sin{\left(\frac{t}{t_0}\right)}-\frac{t}{t_0}\right\},
\end{equation} 
with $\beta=(1-b_{\rm ad})/(2n\pi)$ where $n\geq 1$ is an integer and where the corresponding adiabatic shortcut times are $\tau_n=2n\pi t_0$. For the corresponding protocol obtained via Equation \eqref{Ermarkov2}, $\om(t)$ need not be real (as $\omega(t)^2$ can be negative).
For example, consider $t_0=1/(2\alpha\pi\omega_1)$ where $\alpha>0$ is a positive real number. Then, $\omega(t)^2\geq 0,\ \forall t\in[0,\tau]$ for $2\leq\alpha \leq 2.75$ if $x\equiv \omega_1/\omega_2 \leq 0.1$. Therefore, this protocol can beat the lower bound for the protocol defined by Equations \eqref{accprotAB} and \eqref{accprotCD}, for very small frequency ratio $\omega_1\ll\omega_2$. It is worth mentioning that after exact numerical computation, we find that $\tau_1\leq 1/\omega_1$ for all values of $x$ and is also less than the minimal time for STA found in Equation~\eqref{taun} (obtained with $n=1$). Actually, if one considers $\tau$ and $\omega_{1(2)}$ fixed, Equation \eqref{bSTAex} along with Equation~\eqref{Ermarkov2} define a unique protocol (assuming $\omega(t)^2>0$ for all $t\in [0,\tau]$), with $t_0$ being the solution of the equation $\omega(\tau)=\omega_2$. In this case, we find another instance for which the minimum time for having an STA is smaller than the one required for the existence of an accidental STA described in \mbox{Section \ref{AccSTA};} see~Equation \eqref{taun}.

\subsubsection{Counterdiabatic Driving}\label{secCD}

Demirplak and Rice introduced a formalism to ``run a fast-motion video'' of the adiabatic dynamics of a quantum system by assisting the evolution with counterdiabatic driving fields \cite{Demirplak03}. An equivalent approach was developed independently by Berry \cite{Berry09}. The technique is known as counterdiabatic driving (CD) and also referred to as transitionless quantum driving. It was extended to many-body systems in \cite{DRZ12,delcampo13,DJD14,Saberi14}.
For a single-particle QHE, the use of CD was discussed in \cite{Deng13}.
We~next consider the performance of a many-particle QHE assisted by STA based on CD.

Let the instantaneous eigenvalue problem of the working-medium Hamiltonian with $\om~=~\om(t)$~read:
\beqa
\hat{H}(\om)|n(\om)\ra=E(\om)|n(\om)\ra.
\eeqa

In its original formulation, CD ensures that the adiabatic approximation:
\beqa
\label{adiabsol}
|\psi_n(t)\ra=\exp\left(-i\int_0^tdt'\varepsilon_n(t')-\int_0^tdt'\dot{\om}\la n|\partial_{\om}n\ra \right)|n\ra,
\eeqa
becomes the exact solution of the time-dependent Schr\"odinger equation, when the original Hamiltonian is assisted by the additional term:
\beqa
\label{H1}
\hat{H}_1&=& i\hbar\dot{\om}\sum_n(|\partial_\om n\ra\la n|-\la n|\partial_\om n\ra |n\ra\la n|).
\eeqa
That is, given an initial state the dynamics generated by the CD Hamiltonian $\hat{H}_{\rm CD}=\hat{H}+\hat{H}_1$ is exactly given by the adiabatic approximation to $\hat{H}[\om(t)]$.
CD removes the requirement of slow driving, while the norm of the auxiliary CD term increases with decreasing running time \cite{Demirplak08,DRZ12}.
Further, we note that the auxiliary term generally involves multiple-body interactions whose experimental realization is challenging \cite{DRZ12,Saberi14}.
However, in the presence of scaling dynamics, a remarkable simplification occurs; not only $\hat{H}_1$ can be computed in a closed form without the use of the spectral properties of the Hamiltonian as an input, but it takes the form of a one-body counterdiabatic term \cite{delcampo13}.
For the working medium Equation \eqref{eq:csm}, the driving Hamiltonian is:
\begin{equation}
 \hat{H}_{\rm CD}=\sum_{i=1}^{\N}\left[-\frac{\hbar^2}{2m}\frac{\partial^2}{\partial z_i^2}+\frac{1}{2}m\om(t)^2 z_{i}^{2}+\frac{\dot{b}_{\rm ad}}{2b_{\rm ad}}\{z_i,p_i\}\right]
+\frac{\hbar^2}{m}\sum_{i<j}\frac{\lambda(\lambda-1)}{(z_i-z_j)^2},
 \label{eq:csmSTA2}
\end{equation}
where $p_i$ is the momentum operator canonically conjugated to $z_i$ and $\{z,p\}=zp+pz$ is the anticommutator.
The exact solution of the time-dependent many-body Schr\"odinger equation is then given by:
\begin{equation}
\Psi_{\rm CD}\left(z_1,\dots,z_\N,t\right)=\frac{1}{b_{\rm ad}^{\N/2}}\Psi\left(\frac{z_1}{b_{\rm ad}},\dots,\frac{z_\N}{b_{\rm ad}},t=0\right).
\end{equation}

Under CD, the mean energy equals the adiabatic value $\la\hat{H}_{\rm CD}(t)\ra= \la\hat{H}(0)\ra/b_{\rm ad}^2(t)$ during the whole time-evolution for an arbitrary protocol $\om(t)$, and the nonadiabatic factor reduces to unity, \emph{i.e.}, 
\beqa
Q_{\rm CD}^{\ast}(t)=1,
\eeqa
 as shown in Figure \ref{figureQ}. 
Note that the adiabatic evolution Equation (\ref{adiabsol}) associated with $\hat{H}(\om)$ is nonadiabatic with respect to the instantaneous eigenbasis of the CD Hamiltonian (\ref{eq:csmSTA2}).

\subsubsection{Local Counterdiabatic Driving}\label{secLCD}

The experimental implementation of the squeezing operator $\propto\sum_i\{z_i,p_i\}$ required as a counterdiabatic term discussed in the previous section, understood either as a auxiliary external control or as part of the intrinsic Hamiltonian describing the working medium, is generally a challenging task. As a result, it is natural to look for a unitarily-equivalent Hamiltonian in which the counterdiabatic term is mapped to a local potential. This has been achieved for a large family of many-body systems with scale-invariant dynamics \cite{delcampo13,DJD14}, as well as in the single-particle case \cite{delcampo11}. 
The Hamiltonian (\ref{eq:csm}) belongs to such a family. In this case, the driving of the working medium with a local counterdiabatic control (LCD) reads
\beqa
 \hat{H}_{\rm LCD}=\sum_{i=1}^{\N}\left[-\frac{\hbar^2}{2m}\frac{\partial^2}{\partial z_i^2}+\frac{1}{2}m\Omega^2(t) z_{i}^{2}\right]+\frac{\hbar^2}{m}\sum_{i<j}\frac{\lambda(\lambda-1)}{(z_i-z_j)^2},
 \label{eq:csm2}
\eeqa
with:
\begin{equation}
\label{q64}
\Om^2(t)=\om^2(t)- \frac{\ddot{b}_{\rm ad}}{b_{\rm ad}}=\om^2(t)-\frac{3}{4}\frac{\dot{\om}^2}{\om^2}+\frac{1}{2}\frac{\ddot{\om}}{\om}\, .
\end{equation}

The time-evolving state is then given by
\beqa
\Psi_{\rm LCD}\left(z_1,\dots,z_\N,t\right)
=\exp\left(i\sum_{i=1}^\N\!\!\frac{ m\dot{b}_{\rm ad}}{2\hbar b_{\rm ad}}z_i^2\right)\Psi_{\rm CD}\left(z_1,\dots,z_\N,t\right)\ ,
\eeqa
while the corresponding mean energy is
\beqa
\label{aveSTA3}
\la \hat{H}_{\rm LCD}(t)\ra=\frac{1}{b_{\rm ad}^2}\la \hat{H}(0)\ra+\sum_{i=1}^{\N}\frac{\dot{b}_{\rm ad}}{2b_{\rm ad}}\la\{z_i,p_i\}(0)\ra+\sum_{i=1}^{\N}\frac{m}{2}(\dot{b}_{\rm ad}^2-b_{\rm ad}\ddot{b}_{\rm ad})\la z_i^2(0)\ra.
\eeqa

Assuming the initial state to be at equilibrium so that $\la\{z_i,p_i\}(0)\ra=0$, it follows that
\beqa
\label{aven}
\la \hat{H}_{\rm LCD}(t)\ra=\frac{Q_{\rm LCD}^{\ast}(t)}{b_{\rm ad}^2}\la \hat{H}(0)\ra.
\eeqa

Under LCD, the nonadiabatic factor along the process is given by
\beqa
Q_{\rm LCD}^{\ast}(t)=1+\frac{1}{4}\left(\frac{\ddot{\om}}{\om^3}-\frac{\dot{\om}^2}{\om^4}\right),
\eeqa
and reduces explicitly to unity at the beginning and end of the LCD protocol, provided $\dot{\om}(0)~=~\dot{\om}(\tau)~=~0$. We mention that this equation can also be derived using our general equation \eqref{Qstar} for $Q^\ast$ with the substitutions $b\mapsto b_{\rm ad}$ and $\omega(t)\mapsto \Omega(t)$. 
This is illustrated in Figure \ref{figureQ} for the driving protocol $\om(t)$ varying from $\om_1$--$\om_2$ and satisfying these boundary conditions together with 
$\ddot{\om}(0)=\ddot{\om}(\tau)=0$. The explicit expression for $\om(t)$ resembles that for the scaling factor in Equation \eqref{bpolyn} and reads
\beqa
\omega(t)=\omega_1+\alpha_3t^3+a_4t^4+a_5t^5\ ,
\eeqa
where $\alpha_3=10(\omega_2-\omega_1)/\tau^3$, $\alpha_4=-15(\omega_2-\omega_1)/\tau^4$ and $\alpha_5=6(\omega_2-\omega_1)/\tau^5$.
The nonadiabatic nature of the STA designed by LCD becomes apparent in Figure \ref{figureQ}a, where transient excitations are generated during the protocol and canceled out upon its completion. At variance with accidental STA, LCD protocols can be engineered for arbitrary small values of $\tau$ provided that the modulation of the trap frequency $\Om(t)$ can be implemented; see Figure \ref{figureQ}b.

\section{Discussion}

Along the compression and expansion strokes of a quantum Otto cycle, the nonadiabatic dynamics of a many-particle working medium can be efficiently characterized by the nonadiabatic factor $Q^*(t)$ whenever the dynamics is scale-invariant, as is the case for the family of models described by Hamiltonian (\ref{Hsu11}). 
The adiabatic performance is matched whenever $Q^*(\tau)=1$ at the end of the compression and expansion strokes governed by unitary dynamics. We have seen that the condition $Q^*(\tau)=1$ can be fulfilled without resorting to adiabatic dynamics. In particular, a variety of STA can be exploited as an alternative. This possibility has prompted us to introduce two alternative schemes to boost the performance of a many-particle QHE. In both approaches, the fixed resources are given by the temperature of the hot and cold reservoirs, together with the particle number and the inter-particle interactions.
One can opt for the optimization of the output power of the QHE run in finite-time $\tau$ as a function of the frequency ratio of $\om_1/\om_2$. 
Under the resulting condition for maximum finite-time output power, varying the value of $\tau$ for a given specific functional form of $\om(t)$
 can lead to accidental STA protocols satisfying $Q^*(\tau)=1$, for a set of discrete values $\tau=\tau_n$ ($n=1,2,3,\dots$), as discussed in Section \ref{AccSTA}.
 As an alternative, one can directly opt for maximizing the output power of the QHE assuming zero friction ($Q^*(\tau)=1$) and 
 then use STA in many-particle systems to consistently match the adiabatic performance and reach the maximum efficiency of the cycle. 
 Such STA protocols can be engineered by a variety of techniques, including reverse engineering of the scale-invariant dynamics (Section \ref{secRI}), counterdiabatic driving (Section~\ref{secCD}) and local counterdiabatic driving (Section~\ref{secLCD}).
 Both approaches lead to the operation of the many-particle QHE at maximum efficiency and optimal output power.

\section{Conclusions}

Finite-time thermodynamics aims at optimizing the nonadiabatic performance of thermal machines, required for any realistic application. 
Conditions for optimal performance generally depend on the specific characteristics of the working medium, such as the number of particles and the interaction strength.
At the quantum level, the required optimization involves tailoring thermal and quantum fluctuations.
The maximum efficiency can be reached in the limit of slow driving, at~the expense of a vanishing output power. 

We have proposed to scale up quantum heat engines by using a many-particle working medium in combination with shortcuts to adiabaticity, 
to simultaneously amplify the output power and match the adiabatic performance in a finite cycle time.
The resulting heat engine operates with zero friction at the maximum efficiency and exhibits an enhanced output power as a result of the many-particle nature of the working medium and the reduced cycle time.
The maximum efficiency is then set by the adiabatic Otto upper bound shared by both single- and many-particle heat engines.
This scheme can be implemented realizing a quantum Otto cycle with an interacting trapped Bose or Fermi gas with a suitable modulation of the harmonic trap. 
Our results provide a new avenue for engineering nonadiabatic friction-free scalable thermal machines that should find applications at the interface of quantum thermodynamics and energy science. 

\vspace{6pt} 


\acknowledgments{Funding support from UMass Boston (Project P20150000029279) and European Science Foundation 
~(POLATOM-5052) is further acknowledged.}

\appendix
\section*{\noindent Appendix A. Nonadiabaticity of the Accidental Protocol}
\vspace{6pt}

\renewcommand{\theequation}{A\arabic{equation}}
\setcounter{equation}{0}

Consider the accidental STA protocol defined by Equations \eqref{accprotAB} and \eqref{accprotCD}, with the scaling factor $b(t)$ and nonadiabatic factor $Q^{\ast}_{AB(CD)}$ discussed in Section \ref{AccSTA}. In this Appendix, we compute these quantities for the first step of the Otto cycle $A\rightarrow B$ (compression). One can check that the nonadiabatic factor is symmetric $Q^*_{AB}=Q^*_{CD}$ and that the scaling factor verifies the consistency conditions for a STA for the expansion $C\rightarrow D$, as well as the compression.

\subsection*{Appendix A.1. Derivation of the Scaling Factor $b(t)$}

The scaling factor $b(t)$ is given by the following formula \cite{Pinney,JBdC15}:
\begin{equation}\label{FormulaSF}
b(t)=\sqrt{G_1(t)^2+\omega(0)^2G_2(t)^2},\ \mathrm{for}\ t>0.
\end{equation}

For the sake of simplicity, we put $\phi_t=1-t/t_1$, which leads to $\partial_t\phi_t=-1/t_1$ and $\phi_{t=\tau}=x$. 
The two fundamental solutions are:
\begin{align*}
G_1(t)&= \frac{\left(\phi_t\right)^{\frac{1}{2}}}{2\alpha}\left\{(-1+\alpha)\left(\phi_t\right)^{\frac{\alpha}{2}}+(1+\alpha)\left(\phi_t\right)^{-\frac{\alpha}{2}}\right\},\ \mathrm{for}\ t>0,\\
G_2(t)&= \frac{t_1}{\alpha}\left(\phi_t\right)^{\frac{1}{2}}\left\{\left(\phi_t\right)^{\frac{\alpha}{2}}-\left(\phi_t\right)^{-\frac{\alpha}{2}}\right\},\ \mathrm{for}\ t>0,
\end{align*}
with $\alpha=\sqrt{1-\gamma^2}$ and for $\gamma\neq 1$. 
Given that $1-\tau/t_1=x$, we find:
\begin{align*}
G_1(\tau)&= \frac{\sqrt{x}}{2\alpha}\left\{(-1+\alpha)x^{\frac{\alpha}{2}}+(1+\alpha)x^{-\frac{\alpha}{2}}\right\},\ \mathrm{for}\ t=\tau,\\
G_2(\tau)&= \frac{t_1}{\alpha}\sqrt{x}\left\{x^{\frac{\alpha}{2}}-x^{-\frac{\alpha}{2}}\right\},\ \mathrm{for}\ t=\tau.
\end{align*}

Direct computation yields:
\begin{align*}
G_1(t)^2&=\frac{\phi_t}{4\alpha^2}\left\{(2-\gamma^2)\left(\phi_t^\alpha+\phi_t^{-\alpha}\right)-2\gamma^2-2\alpha\left(\phi_t^\alpha-\phi_t^{-\alpha}\right)\right\},\ \mathrm{for}\ t>0,\\
\omega_1^2G_2(t)^2&=\frac{\gamma^2\phi_t}{4\alpha^2}\left\{\left(\phi_t^\alpha+\phi_t^{-\alpha}\right)-2\right\},\ \mathrm{for}\ t>0.
\end{align*}

Using Formula \eqref{FormulaSF}, we can show that:
\begin{align*}
b(t)^2&=\frac{\phi_t}{2\alpha^2}\left\{-2\gamma^2+\left(\phi_t^\alpha+\phi_t^{-\alpha}\right)-\alpha\left(\phi_t^\alpha-\phi_t^{-\alpha}\right)\right\},\ \mathrm{for}\ t>0,\\
b(\tau)^2&=\frac{x}{2\alpha^2}\left\{-2\gamma^2+\left(x^\alpha+x^{-\alpha}\right)-\alpha\left(x^\alpha-x^{-\alpha}\right)\right\},\ \mathrm{for}\ t=\tau.
\end{align*}

The first derivative of the square of the scaling factor gives:
\begin{align*}
2\dot{b}(t)b(t)&=\gamma^2\frac{\dot{\phi}_t}{2\alpha^2}\left\{-2+\phi_t^\alpha+\phi_t^{-\alpha}\right\},\ \mathrm{for}\ t>0,\\
2\dot{b}(\tau)b(\tau)&=\frac{\gamma^2}{2\alpha^2 x t_1}\left\{2-\left(x^\alpha+x^{-\alpha}\right)\right\},\ \mathrm{for}\ t=\tau.
\end{align*}

Notice that for $\gamma>1$ (\emph{i.e.}, $\alpha=i\sqrt{\gamma^2-1}$), we can rewrite the previous equations by substituting $(2\sqrt{\pm 1})^{-1}(\phi_t^\alpha\pm \phi_t^{-\alpha})$ by $\cos{\left(\sqrt{\gamma^2-1}\ln{(\phi_t)}-\frac{\pi}{4}\pm\frac{\pi}{4}\right)}$.
Using this form, it is easy to show that $b(\tau_n)=1$ and $\dot{b}(\tau_n)=0$, where $n=1,2,3,\dots$ is a positive integer, and the value of $\tau_n$ is given by Equation \eqref{taun}, which sets the STA time. A similar calculation shows that $\ddot{b}(\tau_n)=0$. It is also interesting to note that for small $t$, the Taylor expansion has a similar form to that in Equation \eqref{bpolyn}. 

\subsection*{Appendix A.2. Derivation of the Nonadiabatic Factor $Q^{\ast}$}

To compute the nonadiabatic factor $Q^{\ast}$, we use the Husimi formula \cite{Husimi53}:
\begin{equation}\label{QH}
Q^{\ast}(t)=\frac{\left(\dot{G}_1(t)^2+\omega(t)^2G_1(t)^2\right)+\omega(0)^2\left(\dot{G}_1(t)^2+\omega(t)^2G_1(t)^2\right)}{2\omega(0)\omega(t)},\ \mathrm{for}\ t>0.
\end{equation} 

We note that the square of the first derivative of the fundamental solution of the classical equation reads:
\begin{align*}
\dot{G}_1(t)^2&=
\frac{\gamma^4(\dot{\phi}_t)^2}{16\alpha^2\phi_t}\Big\{\left(\phi_t^\alpha+\phi_t^{-\alpha}\right)-2\Big\},\ \mathrm{for}\ t>0,\\
\omega_1^2\dot{G}_2(t)^2&=\frac{\gamma^4(\dot{\phi}_t)^2}{16\alpha^2\phi_t}\Big\{(2-\gamma^2)\left(\phi_t^\alpha+\phi_t^{-\alpha}\right)-2\gamma^2+2\alpha\left(\phi_t^\alpha-\phi_t^{-\alpha}\right)\Big\},\ \mathrm{for}\ t>0,
\end{align*}
whence it follows that:
\begin{equation*}
\dot{G}_1(\tau)^2+\omega_1^2\dot{G}_2(\tau)^2
=\frac{\omega_1^2}{2x\alpha^2}\left\{-2\gamma^2+\left(x^\alpha+x^{-\alpha}\right)+\alpha\left(x^{\alpha}-x^{-\alpha}\right)\right\},\ \mathrm{for}\ t=\tau.
\end{equation*}

Using Equation \eqref{QH}, we obtain:
\begin{equation*}
Q^{\ast}(\tau)=\frac{1}{2x}b(\tau)^2+\frac{x}{2}\left(\frac{\dot{G}_1(\tau)^2+\omega_1^2\dot{G}_2(\tau)^2}{\omega_1^2}\right),\ \mathrm{for}\ t=\tau,
\end{equation*}
leading to Formula \eqref{QTD}.

\end{document}